\documentstyle[12pt]{article}

\begin{document}
\title{\textbf{Fast magnetic and electric dynamos in flat Klein bottle plasma flows}} \maketitle
{\sl \textbf{L.C. Garcia de Andrade}\newline
Departamento de F\'{\i}sica
Te\'orica-IF-UERJ- RJ, Brasil\\[-3mm]
\vspace{0.01cm} \paragraph*{Recently Shukurov et al [Phys Rev \textbf{E} (2008)] presented a numerical solution of a Moebius strip dynamo flow, to investigate its use in modelling dynamo flows in Perm torus of liquid sodium dynamo experiments. Here, by analogy one presents an electric dynamo on a twisted torus or Klein bottle topology. An exact solution in the form of flat Klein bottle dynamo flow is obtained. It is shown that even in the absence of magnetic dynamos initial electric fields can be amplified in distinct points of the Klein bottle. In this case diffusion is taken as ${\eta}\approx{5.0{\times}10^{-3}{\Omega}-m}$ the electric potential is obtained. The difference of electric fields at the beginning of plasma flow profile is ${\Delta}E_{Dyn}\approx{468\frac{V}{m}}$, which is stronger than the electric dynamo field obtained in the magnetic axis of spheromaks, which is of the order of $E_{Dyn}\approx{200\frac{V}{m}}$. The potential of the dynamo at the surface of the Earth computed by Boozer [Phys Fluids \textbf{B} (1993)] of ${\Phi}\approx{160V}$, is used to obtain the magnetic vector constant $A_{0}\approx{10^{6}V}$. In general the associated magnetic dynamo is shown to be a fast dynamo in force-free plasmas in Klein bottle. Of course if the magnetic dynamo growth is turned on the electric dynamo shall be able to sustain itself along the Klein bottle. PACS: 47.65.Md, 02.40-Ky. Key-words: Dynamo plasmas; Riemannian geometry.}
\newpage
\section{Introduction}
 One of the problems faced by mathematicians and mathematical physicists is to embedded two-dimensional compact Riemannian surfaces \cite{1} in Euclidean three-dimensional ${\textbf{R}}^{3}$ spaces. Though Klein-bottle is one of the flat manifolds that can embedded in ${\textbf{R}}^{4}$ naturally, it does not seem to be possible to embedded isommetrically it in ${\textbf{R}}^{3}$, which can be done trivially with the sphere, cylinder for example. Earlier a flat torus, that can be embedded in ${\textbf{R}}^{3}$ has been used by Arnold et al to build a fast dynamo flow with stretching of magnetic fields on a torus. \newline
 More recently Shukurov et al \cite{2} have provided another example of dynamo flow on a surface that cannot be isometrically embedded in ${\textbf{R}}^{3}$, namely a Moebius strip flow. Examples of simpler embedding of surface dynamo flows in ${\textbf{R}}^{3}$ have been obtained by the author in the context of plasma dynamos as Riemannian magnetic flux tubes \cite{3} in ${\textbf{R}}^{3}$. Actually in this case one has shown that the isommetrically embedding of the Riemannian flux tubes can be embedded in Euclidean 3-manifold. It is well-known \cite{1} that every compact Riemann surface $\cal{M}$ can be endowed with a metric of constant curvature. The interesting example of Shukurov et al \cite{2} though topologically from the Klein bottle, keeps some resemblance with the example given here, since both are isommetrically embedded in ${\textbf{R}}^{4}$. \newline
 Here it is shown that a fast kinematic dynamo \cite{4} with dissipation where the magnetic field differential closed (dB=0) 2-form $\textbf{B}=d\textbf{A}$, where $\textbf{A}=A_{i}dx^{i}$ (i,j=1,2,3), with a constant velocity, is endowed with electric fields which can be comparable to the ones obtained in plasma laboratory in initial stages of the experiment on plasma dynamos, as the one performed by Wang et al \cite{5} using cylinders. Since the cylinder can be locally, considered as a section of a Klein bottle one uses here the data obtained by Wang et al \cite{5} to estimate, the electric field and electric potential. To determine the electric field in the Klein bottle plasma one compares the dynamo action here with the one obtained at the surface of the Earth (geodynamo) by Boozer \cite{6}.\newline
 This paper is organised as follows: Section II addresses the problem of determining the electric dynamo fields and estimate their values in spheromak plasmas using the corresponding data. Section III shows that the magnetic dynamo in the Klein bottle is a fast dynamo, which could not be determined in the numerical solution obtained by Shukurov et al for the Moebius band dynamo action. Discussions are presented in section IV.
\newpage
\section{Electric fields amplification in flat Klein bottle force-free plasmas}
In this section it is shown that the topology of Klein bottle flows is suitable to amplification of electric field dynamos in this plasma flowing topology. Trivial cylindrical geometry constraining plasma dynamos flows have been presented recently by Wang et al \cite{5} which obtained a slow magnetic dynamo. Cylindrical topology is trivially embedded in ${\textbf{R}}^{3}$. In this section, one shall be concerned with the less trivial Klein bottle topology which can be embedded in ${\textbf{R}}^{4}$ whose Riemannian metric is
\begin{equation}
ds^{2}=dx^{2}+(1+3cos^{2}y)dy^{2}
\label{1}
\end{equation}
which is obtained by the following reasoning. The embedding of the Klein bottle in $\textbf{R}^{4}$ is obtained by the definition of coordinates
\begin{equation}
X_{1}=cosy cosx
\label{2}
\end{equation}
\begin{equation}
X_{2}=cosy sinx
\label{3}
\end{equation}
\begin{equation}
X_{3}=2cos\frac{x}{2}siny
\label{4}
\end{equation}
\begin{equation}
X_{4}=2sin\frac{x}{2}siny
\label{5}
\end{equation}
Flat Klein bottle is an analytical map ${\textbf{R}}^{2}/\approx{{\textbf{R}}^{4}}$ \cite{1}, such as that the self-intersection curve is
\begin{equation}
\{{X_{1}}^{2}+{X_{2}}^{2}=1, X_{3}=X_{4}=0\}
\label{6}
\end{equation}
Substitution of the expressions $(\ref{2})-(\ref{4})$ into the flat metric Riemannian line element ${\textbf{R}}^{4}$ as
\begin{equation}
ds^{2}={dX_{1}}^{2}+{dX_{2}}^{2}+{dX_{3}}^{2}+{dX_{4}}^{2}
\label{7}
\end{equation}
reproduces the Riemann-flat metric line element (\ref{1}). At topological intersections $y=0$ and $y=\pi$ one shall find below the values of dynamo electric fields. The electric potential in Klein bottle flow is given by the following gauge
\begin{equation}
{\Phi}={\eta}{\nabla}.\textbf{A}=\frac{{\eta}}{\sqrt{g}}{\partial}_{y}(\sqrt{g}A^{y}(x))
\label{8}
\end{equation}
where $\textbf{A}$ is the vector magnetic potential of the electromagnetic Maxwell theory, and the diffusion constant ${\eta}$ which vanishes in ideal plasmas. Here to simplify matters one chooses $A^{x}$ to vanish. This would leave us with the two-form closed differential form
\begin{equation}
F=B_{z}dx{\wedge}dy
\label{9}
\end{equation}
One notes immeadiatly from this expression that in the electric plasma dynamos, only exists in resistive non-ideal plasmas. By expressing the gradient operator ${\nabla}$ over the Klein bottle as
\begin{equation}
{\nabla}=\textbf{x}{\partial}_{x}+\textbf{y}[1+3cos^{2}y]^{-1}{\partial}_{y}
\label{10}
\end{equation}
By taking the determinant of the Klein bottle metric $g=(1+3cos^{2}y)$ one obtains the following electric dynamo potential
\begin{equation}
{\Phi}_{Dyn}=\frac{3}{2}{\eta}\frac{sin2y}{(1+3cos^{2}y)}A^{y}
\label{11}
\end{equation}
Therefore to completly determines the expression for the electric potential dynamo, one needs to first obtain $A^{y}(x,t)$ vector magnetic potential function, and this is exactly obtained from the dynamo equation
\begin{equation}
\frac{{\partial}\textbf{A}}{{\partial}t}+{\eta}{\nabla}^{2}\textbf{A}=\textbf{U}\times{\nabla}
\textbf{A}\label{12}
\end{equation}
where $\textbf{A}$ is the magnetic vector potential whose magnetic field is given by
\begin{equation}
\textbf{B}={\nabla}\times\textbf{A}\label{13}
\end{equation}
Since the frame of reference in this case is Euclidean this equation reduces to the following component equation
\begin{equation}
\frac{{\partial}{A^{i}}}{{\partial}t}+{\eta}{\nabla}^{2}{A^{i}}=[V_{j}(A^{i,j}-A^{j,i})]
\label{14}
\end{equation}
where the flow velocity $\textbf{V}$ satisfies the solenoidal condition $div\textbf{V}=0$ and it is defined by the relation
\begin{equation}
\textbf{V}=V_{0}\frac{{\partial}}{{\partial}x}
\label{15}
\end{equation}
where $V_{0}$ is constant. Solution of the dynamo equation taking into account the force-free equation and the force-free gauged equation \cite{7}
\begin{equation}
\textbf{B}={\lambda}(\textbf{A}+{\nabla}{\Psi})
\label{16}
\end{equation}
This solution yields
\begin{equation}
{A^{y}}(x,t)=A_{0}e^{{\gamma}(t+\frac{x}{V_{0}})}
\label{17}
\end{equation}
Substitution of this expression into the potential equation above yields
\begin{equation}
{\Phi}_{Dyn}(x,y,t)=\frac{3}{2}{\eta}\frac{sin2y}{(1+3cos^{2}y)}A_{0}e^{{\gamma}(t+\frac{x}{V_{0}})}
\label{18}
\end{equation}
The electric field dynamo becomes
\begin{equation}
\textbf{E}_{Dyn}=-{\nabla}{\Phi}_{Dyn}(x,y,t)
\label{19}
\end{equation}
Which yields the following expression for the electric dynamo field
\begin{equation}
\textbf{E}_{Dyn}=-\frac{3}{2}A_{0}{\eta}[\frac{{\gamma}}{V_{0}}sin2y\textbf{x}+
\frac{[2cos2y+6cos^{2}ycos2y-sin2y]}{(1+3cos^{2}y)^{3}}\textbf{y}]e^{{\gamma}t+kx}
\label{20}
\end{equation}
One immeadiatly notes that in the absence of diffusion the electric fields does not survive in ideal plasmas. It also shows that one may be able to compute the electric dynamo, which is similar to the electric dynamo plasma definition
\begin{equation}
\textbf{E}_{Dyn}={\eta}\textbf{J}
\label{21}
\end{equation}
where $\textbf{J}$ is the electric current in the plasma flow, to obtain its value in two distinct points of the Klein bottle, namely at $(0,0,0)$ origin and the point $(0,\frac{\pi}{4},0)$. These yields respectively
\begin{equation}
\textbf{E}_{Dyn}(0,0,0)=-\frac{3}{16}A_{0}{\eta}\textbf{y}
\label{22}
\end{equation}
By making use of a typical spheromak diffusion constant of ${\eta}\approx{5.0{\times}10^{-3}{\Omega}-m}$ one obtains
\begin{equation}
|\textbf{E}_{Dyn}|(0,0,0)=\frac{15}{16}A_{0}{\times}10^{-3}
\label{23}
\end{equation}
To determine the vector magnetic potential constant ${A}_{0}$, since it is a constant one may apply it to any electric potential calculated previously, as the one by Boozer \cite{6} which yields for the Earth geodynamo an electric potential of $160 V$. Applying this to the above expression for the electric potential yields
\begin{equation}
A_{0}\approx{1.4{\times}10^{6}V}
\label{24}
\end{equation}
which yields
\begin{equation}
|\textbf{E}_{Dyn}|(0,0,0)=132\frac{V}{m}
\label{25}
\end{equation}
Let us now, compare the behaviour of the electric fields in two points of the flat Klein bottle, by computing the value of the electric dynamo field in other point as
\begin{equation}
|\textbf{E}_{Dyn}|(0,\frac{\pi}{4},0)= 600\frac{V}{m}
\label{26}
\end{equation}
which yields the difference between the electric dynamo fields at to points of the Klein bottle flow as
\begin{equation}
{\Delta}|\textbf{E}_{Dyn}|= 468 \frac{V}{m}
\label{27}
\end{equation}
which is more than the double of the electric dynamo field obtained in the magnetic axis of the spheromak which is $E_{Dyn}\approx{200\frac{V}{m}}$. This shows that since both points of the Klein bottle would be continued to be feed by the time exponential growth rate ${\gamma}$, in both points of the Klein bottle, the dynamo action shall be granted and the electric fields growth shall be sustained. As shall be demonstrated in the next section the magnetic dynamo in the Klein bottle can be a fast dynamo under certain conditions on the flow. This is basically if the sign of the ave number is the same as the velocity of the flow.
\newpage
\section{Force-free fast magnetic dynamo in Klein bottle flows} In this section the type of amplification of the magnetic field shall be checked from the self-induction equation above, by examining if the dynamo is either fast or slow. The magnetic field solution shows clearly that the magnetic field growth rate is positive if there is a a positive relation between the wavelength of dynamo and the velocity of the flow. The magnetic self-induction equation
\begin{equation}
\frac{{\partial}\textbf{B}}{{\partial}t}+{\eta}{\nabla}^{2}\textbf{B}={\nabla}{\times}(\textbf{U}\times{\nabla}
\textbf{B})\label{28}
\end{equation}
By solving the equation for the ansatz $\textbf{B}=\textbf{B}_{0}e^{{\gamma}t+kx}$ ($\textbf{B}_{0}=constant$) one obtains the following relation
\begin{equation}
{\gamma}=kV_{0}-{\eta}{\lambda}^{2}\label{29}
\end{equation}
Note that k is magnetic dynamo wavelength. Thus in the fast dynamo limit \cite{4}
\begin{equation}
lim_{{\eta}\rightarrow{0}}\textbf{Re}{\gamma}(\eta)\ge{0}
\label{30}
\end{equation}
The zero sign of course represents the marginal dynamo and $\textbf{Re}$ represents the real part of the growth rate ${\gamma}$, when it applies. Thus note that the dynamo is fast if the following relation is obeyed
\begin{equation}
kV_{0}\ge{0}
\label{31}
\end{equation}
Since here ${\eta}\approx{LV_{0}Rm^{-1}}$ is the diffusion relation with the magnetic Reynolds number $Rm$, the relation (\ref{29}) becomes
\begin{equation}
{\gamma}=V_{0}[\frac{2{\pi}}{{\lambda}_{0}}-L{Rm}^{-1}{\lambda}^{2}]\label{32}
\end{equation}
where ${\lambda}_{0}$ is the dynamo wavenumber. Therefore, fast dynamo action in the Klein bottle implies the following constraint between the dynamo wavenumber and the magnetic Reynolds number as
\begin{equation}
\frac{{\lambda}_{0}}{2{\pi}}\ge{{Rm}L^{-1}{\lambda}^{-2}}\label{33}
\end{equation}
when the flow obeys the constraint $V_{0}\ge{0}$. Note that from this last expression, the plasma dynamo laboratory date obtained by Wang et al \cite{6} of $Rm=210$ and the typical length of a plasma experiment $L=1m$ for a growth rate of ${\gamma}\approx{0.022s^{-1}}$ yields the following plasma dynamo wavelength ${\lambda}_{0}$ as
\begin{equation}
\frac{{\lambda}_{0}}{2{\pi}}=L^{-1}{Rm}{\lambda}^{-2}+{\gamma}\label{34}
\end{equation}
and numerically estimated as
\begin{equation}
\frac{{\lambda}_{0}}{2{\pi}}=210{\lambda}^{-2}+0.022\label{35}
\end{equation}
which shows clearly that the coupling force-free dynamo constant ${\lambda}$ being very small here by assumption leads us to a long dynamo wave-length or large-scale dynamo. Let us now use the following Riemann-Christoffel ${{\Gamma}^{i}}_{jk}$ only surviving component in Klein bottle as
\begin{equation}
{{\Gamma}^{y}}_{yy}=\frac{3}{2(1+3cos^{2}y)}sin2y
\label{36}
\end{equation}
where the Riemann-Christoffel connection
${{\Gamma}^{i}}_{jk}$ only surviving component in Klein bottle is
\begin{equation}
{{\Gamma}^{i}}_{jk}=\frac{1}{2}g^{il}({g_{lj}}_{,k}+{g_{lk}}_{,j}-{g_{jk}}_{,l})
\label{37}
\end{equation}
The non-geodesic equation is
\begin{equation}
\frac{dV^{y}}{ds}+{{\Gamma}^{y}}_{y}(V^{y})^{2}=F^{y}(y)
\label{38}
\end{equation}
Since $V^{y}$ is constant this equation reduces to
\begin{equation}
{{\Gamma}^{y}}_{y}(V^{y})^{2}=F^{y}(y)
\label{39}
\end{equation}
or explicitly written, the dynamo force is
\begin{equation}
F(y)=\frac{3{V_{0}}^{2}}{2(1+3cos^{2}y)}sin2y
\label{40}
\end{equation}
At the point $(0,0,0)$ the force vanishes and at $(0,\frac{\pi}{4},0)$ this yields the following force
\begin{equation}
F(\frac{\pi}{4})=\frac{3}{5}{V^{2}}_{0}
\label{40}
\end{equation}
Note that this force is a nonlinear viscous force in the Klein bottle. Actually the external force able to sustain the dynamo action is periodic and bounded. Actually this is a Coriolis force, due to the action of the frame used, and is similar to the one used by Einstein in his general theory of relativity.

\section{Conclusions}
One of the problems facing theoretical people working in dynamo theory is to decide whether the dynamo is either fast or slow. Recently Shukurov et al have been working on a numerical solution which did not helped them to decide if the dynamo as fast or slow, since they did not obtain analytical solutions in the Moebius strip dynamos. This is actually one of the main motivations for the study undertaken here, here analytical solutions yields fast magnetic dynamos in Klein bottle flows, and electric dynamos. In this paper one makes use of an external force model on kinematic fast dynamos in 2D Klein bottle manifolds of zero Riemannian curvature.
\section{Acknowledgements}
Several discussions with A Brandenburg, D Sokoloff and Yu Latushkin are highly appreciated. I also thank financial  supports from UERJ and CNPq. Financial supports from both institutions are greatful acknowledged.
 
  \end{document}